\documentstyle[eqsecnum,aps]{revtex}
\begin{document}

\title{Quantum Hall Transitions in Field Induced Spin
 Density Wave Systems}
\author{Pascal Lederer and Baruch Horovitz$^*$}
\address{Physique des Solides, Universit\'e Paris-Sud, 91405-Orsay cedex, 
France (laboratoire associ\'e au C.N. R. S.), $^*$ On leave  from
Department of Physics, Ben-Gurion University, Beer-Sheva, Isra\"el.}

\maketitle

\begin{abstract}
Field induced spin density wave (FISDW) systems exhibit coexistence
phases between well defined quantum Hall plateaux phases 
with even integers $2N$ and 
$2N'$ . We show that a disordered coexistence region accounts for the observed 
peaks in the longitudinal reisitivity as the field varies between plateaux.
 It also results in a random spin
mixing which yields two energy split extended states. The longitudinal
resistance is expected to show two peaks with a temperature ($T$) dependent 
width $\sim T^\kappa$. The peak width should saturate below the non-nesting
interlayer coupling of $\approx 40 mK$.
\end{abstract}
D.Electronic transport; Quantum Hall effect
\vspace{20mm}
 
 Organic conductors  exhibit  a cascade of (Magnetic) Field Induced 
 Spin Density Wave Phases  (FISDW) below about 1K and in fields 
 ranging from a few Teslas 
 to about 20 Teslas. Each SDW phase
  shows a well defined Hall
 plateau where the Hall resistance is $h/(2e ^2N)$ with an integer
 $N$ \cite{review}. The sequence of integers is usually monotonic, 
 the integer $N$ decreasing by 1 as the field increases.
  The even integers $2N$ signify that both spin states are coupled
 by the spin density wave (SDW) and the quantum Hall (QH) phenomena 
 is then degenerate in
 the spin states.
  Although the Quantized Hall Effect  (QHE) in FISDW seems similar
   in many respects to the
  Integer QHE seen, e.g., in MOSFETS (\cite{prange}), it is different
from the latter in many important ways: it is observed in an anisotropic three 
dimensionnal material; the effect would not exist in the absence of 
electron-electron 
interactions; furthermore, under specific conditions of pressure 
and field, the QHE changes sign:
a negative plateau  is inserted within the positive sequence of FISDW 
\cite{balicas}; last, disorder plays no role in the phenomenon, 
contrary to the situation
in the IQHE: the Fermi level is pinned in the middle of the SDW 
electronic gap between
 extended states by the broken symmetry phenomenon, not by disorder.
  In fact the very existence of
 FISDW phases is made possible because very clean samples are 
 available. This  is a prerequisite 
 for the observation of FISDW: the electronic mean free 
 path has to be larger than the 
 magnetic length $2\pi/G=\phi_0/(Bb) $ (where $\phi_0$ is the 
 flux quantum, $B $ 
 the magnetic field and $b$ the interchain distance in 
 the most conducting plane).

The FISDW phase diagram is well understood within the so called Quantum
 Nesting Model \cite{review,gl,hml}: in materials with open Fermi surfaces 
 and good nesting properties, 
 electronic motion under magnetic field becomes one dimensional and periodic; this 
 opens up gaps between Landau bands when electron-electron interactions stabilize a SDW 
 phase. Minute changes in the electronic dispersion relation, such as can be caused by
  applying pressure, may result in stabilizing phases with negative Hall numbers
  \cite{zanchi}.
  
 The  change in order parameter from sub-phase to sub-phase in the 
 sequence of FISDW is a discontinuous jump of the SDW wave vector parallel 
 component by one inverse magnetic length  G (in the usual monotonic sequence) or by
 an integer number of G's (in the case of a transition from a positive Hall number
 to a negative one). As a result, all phase transitions are weakly first order.
 Hysteresis is observed quite generally both in transport properties
 \cite{balicas} and thermodynamic ones\cite{review}.
   
   Although the Quantum Nesting Model accounts in a satisfactory way for
 nearly all aspects of the phase diagram (except perhaps at very large fields),
 for the existence of Hall plateaux, their sequence, and the rare occurrence of
 negative Quantum Hall numbers, it fails to account correctly for the 
 longitudinal Hall resistance at the transition between FISDW phases. A naive
 interpretation of the model would predict a small discontinuous change 
 of $\rho_{xx}$ at the transition between plateaux at low temperatures,
 reflecting the small discontinuous change in electronic gap at the Fermi level,
 and the activated nature of dissipation processes. The experimental
 situation is quite different: $\rho_{xx}$ exhibits spikes at the 
 transition between plateaux\cite{balicas}, in a manner similar, at first sight,
 to what is observed in the usual MOSFET IQHE\cite{prange}.

 The purpose of this paper is to suggest that spikes
  in $\rho_{xx}$ can be understood 
 on the basis of the thermodynamics of the first order transition between 
 FISDW: we consider the coexistence region between
Hall plateaux $N$ and $N'$ and show that it exhibits critical
phenomena specific to these QH states.
 We propose then the following
scenario. The first order transition between plateaux $N$ and $N'$ is
driven by a balance of the interaction between electrons and Landau
quantization in a magnetic field. Within the coexistence region we
assume that nucleation of the new phase is a slow dynamic process and
therefore the system is composed of isolated clusters of the $N'$ phase
embedded in a continuum of the $N$ phase. On the time scale of electronic
transport, these clusters are randomly quenched. Uncoupled chiral gapless states
are formed at the cluster boundaries.  As the field increases the
coexisting phases have an increased fraction of the $N'$ phase until
a percolation threshold is achieved at some field, beyond which the
$N$ phase forms isolated clusters within the $N'$ phase.

The  finite size of an $N$ cluster
implies that not all the electronic states participate in forming the SDW,
i.e. there should be $2N$ gapless edge states, as in the usual QH
system. In fact Yakovenko and Goan \cite{review} have explicitely
constructed such edge states for the FISDW. They show that the SDW
couples opposite $\pm k_F$ states which are $N$ chains apart where
$k_F$ is the Fermi wave-vector along the quasi one-dimensional chains
($x$ direction). Hence
the last $N$ chains of either $+k_F$ or $-k_F$ near the edge are
uncoupled, i.e. gapless chiral electronic 1D liquids. 
When an $N$ phase is embedded in an
$N'$ phase the latter produces $2N'$ edge states with opposite chirality
around the $N$ phase. The SDW can then couple some of these states, so
that only $|N-N'|$ chains remain gapless. The density of states of the
gapless states is then $N(0)= |N-N'|/(\pi v_F l)$ per unit area, where $l$ 
is the transverse size of a cluster.

Transport in the coexistence region is determined by the gapless
states which are scattered by the random SDW clusters. The problem is then
 similar to the two dimensional (2D) QH
systems in which two spin states are randomly coupled, e.g. by
spin orbit coupling. Here, the two spin states are coupled by 
the SDW fluctuating field. This coupling leads to two nondegenerate extended
states \cite{Ko,KHA1},  which yield peaks in  $\rho_{xx}$; the temperature
dependence of these peaks has been extensively studied in the 2D QH
systems \cite{kochwei} and yields information on the criticality
of QH states  \cite{Polyakov,KHA2}.

Consider the Hamiltonian in presence of a SDW order parameter with
amplitude $\Delta$ and phase $\theta$. This Hamiltonian has been
applied successfully for the FISDW phases \cite{review}.
 We represent the Hamitonian in
a spinor state
\begin{equation}
[u(x,y)\exp (ik_Fx+ik_zz), v(x,y)\exp
(-ik_Fx+i(k_z+\pi /c)]
\end{equation}
for the right and left moving electrons. In all the FISDW phases the SDW has
wavevectors  $\pi /c$ in the least conducting $z$ direction parallel to
the magnetic field \cite{hml}; hence the SDW couples states
with momentum $k_z$  only to those with momenta $k_z+\pi  /c$. This is
a result of perfect nesting in the $z$ direction. i.e. the one-electron
dispersion relation along $k_z$ is dominated by a $t_c \cos(k_zc)$ term
 which allows perfect
matching of the opposite Fermi surfaces when $k_z$ is shifted by
$\pi/c$ \cite{hml,HGW}. In
contrast, the SDW wavevector components $Q_x,Q_y$ depend on the
magnetic field and jump discontinuously between the $N$ phases. In
particular $Q_x =2k_F - NG$.
The Hamiltonian, with Pauli matrices $\tau _i,
i=1,2,3$ in this spinor space, has the form \cite{review}

\begin{equation}
{\cal H}=\{-iv_F \partial _x  - t_c \cos(k_z c)\}\tau_3
+\Delta \tau _1 \exp[i\tau _3 (NG x-\theta)] -f(k_y b-G x)
\end{equation}
Here $v_F$ is the Fermi velocity, and $f(k_yb)$ represents the
electron dispersion in the $k_y$ direction with the wavevector $k_y$ shifted
by a vector potential. Note that the $t_c\cos k_zc$ form is essential in 
obtaining a coupling between the two states in the spinor 
 Eq. (1) and  $t_c\cos [(k_z+\pi)c]=-t_c\cos k_zc$ is used to obtain
 its $\tau_3$
 form in Eq. (2). As a result, by a unitary
transformation $U=\exp [ixt_c\cos (k_zc)/v_F ]$ the $t_c$ term can be 
eliminated and has no effect on the mean field level \cite{gl,hml,HGW}. Non nesting 
terms, e.g. $t'_c\cos (2k_zc)$, cannot be simultaneously transformed away.

In the coexistence phase the SDW order corresponds to a random mixture
of $N$ and $N'$ phases and $\Delta, \theta$ become space dependent. We
assume first that the disorder depends only on $x,y$, i.e. the clusters
are correlated in the $z$ direction. This is reasonable since
as discussed above, variations in
$Q_x$ induce variations in $Q_y$ but not in $Q_z=\pi/c$. Hence
disorder in the $x,y$ directions is an inherent feature of the
coexistence, while the clusters can remain correlated in the $z$
direction. The last term in the Hamiltonian is then $f(-i\partial_y b-G x)$.

We proceed to describe the localization properties of the gapless
states in presence of a random distribution of the two 
coexisting SDW order parameters.
 We recall first the
description of (spinless) electrons in random 2D
QH systems. The transition between QH plateaux is a quantum
percolation transition which involves tunneling and
interference between clusters. It leads to a well known localization
length $\xi$ which diverges as the electronic Fermi energy $E$ approaches a
percolating value $E_c$ as $\xi \sim |E-E_c|^{-\nu}$ with
 $\nu \approx 2.4$ (close
to $7/3$)\cite{prange,Ko}. Quantum percolation signifies here the presence of an
extended state, while at energies away from percolation the states are
localized.
 Note, however, that in the QH system the clusters result
from a given random potential while in the SDW system the disordered
clusters themselves are generated by the magnetic field which drives the 1st
order transition. In both cases the magnetic field drives the Fermi
energy across a percolation point and the dynamics of gapless modes in
the SDW system is similar to that of electrons in a random potential.
In a layered QH system with weak hopping $t$ between layers, an
 uncorrelated disorder between layers \cite{CD}leads to a finite
  width of extended states (scaling as $(t)^{1/ \nu}$) and the actual 
  localization exponent becomes $ \nu \approx 1.45$.
  
The case with two spin states requires some care in identifying their
symmetry.
E.g. if $f(k_yb)=t_b \cos (k_y b)$, i.e. perfect nesting also in the y
direction, the Hamitonian would anticommute with
 $K=\tau _2 \exp (-i\tau _3 \theta)$ leading to particle-hole symmetry. Such
symmetries are essential for identifying universality classes of QH
systems. E.g., in a random superconductor \cite{KHAC} an electron-hole
symmetry is exact and leads to degenerate extended states and distinct
critical exponents; the
symmetry operation in the latter case is antilinear, while the SDW
type (approximate) symmetry defines yet another symmetry class
\cite{Hikami}. Although this SDW symmetry is approximate in the
uniform SDW state, it breaks down in the random coexisting phase. The
phase $\theta$ which signifies an SDW translation is now randomly
$x,y$ dependent
and the operator $K=\tau _2 \exp [-i\tau _3 \theta(x,y)]$ no longer
anti-commutes with the Hamiltonian. The coexistence phase is then
identified as a U(2) symmetry class \cite{Ko,KHA1} in
which the extended states are nondegenerate and have the usual
exponent $\nu \approx 2.4$.

Consider now the localization problem of gapless states in the
presence of a random SDW which mixes the two spin states. We propose
that this is
equivalent to the QH $U(2)$ system \cite{Ko,KHA1} 
with a random scalar potential and 
 a random spin-flip coupling which mixes the two spin states. The
latter system exhibits ``repulsion'' between extended states,
i.e. even if the spin states were degenerate (i.e. no Zeeman term) the
$U(2)$ mixing produces two non degenerate energies of extended
states. A Zeeman term will further increase the splitting. In the SDW
problem the magnetic field drives also the ``landscape'' of the random
potential so that the splitting between critical field values
corresponds to situation that the SDW fluctuation $<
 (\delta \Delta)^2 > ^{1/2}$ changes by $\Delta E$, the
energy splitting of the two extended states. Since fluctuations
$\delta \Delta$ relate to
the unknown kinetics of the first order transition,
 we cannot estimate the splitting of the
fields $H_c$.

 We can, however, evaluate the critical behavior near one
$H_c$ since the situation there is equivalent to a ``spinless''
particle localization. 
As the field approaches a percolation point $H_c$ the localization
length diverges as $\xi \sim |H-H_c|^{-\nu}$ and the resulting
extended state will produce a peak in $\rho_{xx}$. To estimate the
width of this peak at finite temperatures we consider the states at
half maximum of $\rho_{xx}$ as localized states and evaluate their
conductance via variable range hopping, similar to the QH treatment
\cite{Polyakov,KHA2}. The excitation energy for a hop is either
a Coulomb energy or determined by the level spacing of the edge
states. In view of the huge dielectric constant of the SDW state
($\sim 10^9$ or $\sim 10^3$ in the $x$ and $y$ directions,
respectively \cite{dielectric}) we consider an excitation energy which
is dominated by the level spacing $\approx 1/N(0)r^2$ at distance
$r$. Hence
\begin{equation}
\rho_{xx} \sim \exp [\frac{1}{N(0)r^2T}-\frac{r}{\xi}] .
\end{equation}
Minimizing with respect to r yields 

 $\ln \rho _{xx} \sim [(H-H_c)^{2\nu}/N(0)T]^{1/3}$

i.e. for a constant $N(0)$ the width of $\rho_{xx}$ is $|H-H_c|\sim
T^{1/2\nu}=T^{\kappa}$, so that $\kappa=1/(2\nu)$. Hence $\kappa=0.21$ 
for correlated disorder in the $c$ direction while $\kappa=0.34$ for 
uncorrelated disorder.
 Furthermore at a given $H\ne H_c$ this
predicts the Mott law $\ln \rho _{xx} \sim T^{-1/3}$. (Note that, in contrast,
when Coulomb interactions dominate electron-hole levels, 
$\kappa=1/\nu$\cite{Polyakov,KHA2}). 

This derivation
assumes that $N(0)$ is non-critical and smooth near percolation,
 i.e. $N(0)$ of the edge states is
determined by the coexisting clusters rather than by the percolation
path that the edge states choose to take.
Indeed, $N(0)$ is non critical in the usual QH systems \cite{prange,Ko}
 An alternative derivation of the temperature scaling is based on 
 limiting $\xi$ by an inelastic length $L_{\phi}\sim T^{p/2}$\cite{kochwei,wang}
 i. e.  the width of $\rho_{xx}$ is$\sim (L_{\phi})^{-1/\nu}\sim T^{p/2\nu}$
 with $p/2\nu \simeq 0.36$ for the single layer system \cite{wang}.

We consider finally the issue of interlayer coupling. Within our 2D
random system the effective 2D coupling is the deviation from nesting
in the $z$ direction which is rather small $40 mK$ \cite{mhl}. Thus the $\rho
_{xx}$ width should saturate below $40 $ mK.
 If, however, the clusters are not correlated in the $c$ direction, then
  $t_c\approx 10 $K would lead to a large width for  $\rho_{xx}$ and 
  critical behaviour below $\approx 10$ K could not be seen. Recent 
  data on $\rho_{zz}$ \cite{moser} has shown no special features at the 
  transition between plateaux in the $ClO_4$ salt. This is consistent with
   correlated clusters in the $z$ direction. In this case the 
   hopping term $t_c$ in Eq. (2) can be eliminated and $\sigma_{zz}$ depends
    only on the much smaller $t'_c$. In the presence of  random point impurities
    \cite{balents} $t_c$ cannot be strictly gauged a	way, though for extremely
     clean samples either a renormalized $t_c$ or $t'_c$ are the relevant scale.

The explanation  we have given for the behaviour
of the longitudinal resistivity at the transition between FISDW phases is 
based on the first order nature of this transition and on the consideration
of chiral edge electronic liquid states which should exist at the boundaries
of a finite cluster of phase $N$ embedded in a sea of phase $N'$. Our analysis
has various consequences: the intensity of the longitudinal resistivity 
spike at its maximum 
at the transition between plateaux of a monotonic sequence, i.e. when $N-N'=1$
should not depend on field, only on temperature, since it only depends on the 
number of percolating channels at the transition. It should be much larger at
 the transition between phases with different signs of the Hall effect, since 
 the number of dissipative edge channels is $|N-N'|>>1$. In fact the width of
  the $\rho_{xx}$ peak, following eq. (3) has a factor $\delta H\propto [N(0)]^
  {1/2\nu}\propto |N-N'|^{\kappa}$, i.e. increasing with $|N-N'|$.

  Indeed, experiments
 exhibit much larger spikes at the transition between phase $N=-2$ and its 
 neighbours (i.e. with $|N-N'|=4$ or $5$) 
 than between phases for which $N-N'=1$ \cite{balicas}.

  In $(TMTSF)_2PF_6$, a single spike is observed  at all transitions, indicating that
 the spin splitting we expect is too small to be resolved. However
 magnetocaloric and partial transport  data on $(TMTSF)_2 ClO_4$ 
 \cite{Chaikin} shows a
 splitting of the transitions above 4.5 T. Those splittings have been explained 
 phenomenologically on the basis of a Landau Ginzburg expansion and the 
 hypothesis of a repulsive coupling between neighbouring phases.
 A different possible interpretation of this 
 splitting is the spin effect we have discussed above. Higher sensitivity
 data  on $\rho_{xx}$ are required to test our interpretation. However one 
 should keep in mind that the physics of the $ClO_4$ salt is made complicated by 
 the anion ordering problem.

The explanation we have given for the behaviour of the longitudinal resistivity at the
transition between Ultra Quantum Crystal phases (i. e. transition 
between FISDW subphases \cite{lp}) resolves a long standing
problem in the "standard model" ( i.e. Quantum Nesting Model) approach to the Ultra
Quantum Crystal phenomenon. It  is based on the first order
nature of this transition and on the consideration of chiral edge electronic liquid
states which should exist at the boundaries of a finite cluster of phase $N$ embedded
in a sea of phase $N'$. The coexistence regions of the phase diagram, characterized by a
random mixture of clusters of one phase embedded in the other are the regions where
$\rho_{xy}$ varies rapidly with field from one Quantum Hall plateau to the other; we
have shown how this leads naturally to dissipation and to peaks in $\rho_{xx}$,
 where  disorder is induced only by the phase coexistence at the 
 first order transition. Using a variable range hopping approach, 
 we have derived critical exponents for the width of the
$\rho_{xx}$ peaks  and  predict spin splitting of these  peaks.

 An intriguing 
possibility is that the splitted transition observed in the $ClO_4$ salt corresponds to the
resolved spin splitting limit. More detailed experimental data are needed to check the
validity of our critical exponents predictions.

Acknowledgements

B.H. thanks the Arc En Ciel fellowship and the Laboratoire de Physique des Solides for 
hospitality and support. 
We thank C. Caroli, H. S. Goan, D. J\'er\^ome, G. Montambaux, J. Moser, 
and  B. Roulet  for stimulating discussions.

\end{document}